\documentclass[preprint,authoryear,12pt]{aastex}

\usepackage{epsfig}

\usepackage{amssymb}
\usepackage{amsbsy}
\usepackage{subfigure}
\usepackage[mathcal]{euscript}
\usepackage{float}
\usepackage{graphicx}

\bibpunct[]{(}{)}{;}{a}{}{,}

\begin{document}
\slugcomment{Submitted to The Astrophysical Journal Letters}
\title{Implications of Rapid Core Rotation in Red Giants for Internal Angular Momentum Transport in Stars}
\shorttitle{Implications of Rapid Giant Core Rotation}
\author{Jamie Tayar\footnotemark[1] \ and Marc H. Pinsonneault}
\affil{Department of Astronomy, The Ohio State University, 140 W 18th Avenue, Columbus, OH 43210, USA}
\footnotetext[1]{tayar.1@osu.edu}
\shortauthors{Tayar \& Pinsonneault}
\begin{abstract}
Core rotation rates have been measured for red giant stars using asteroseismology. This data, along
with helioseismic measurements and open cluster spin down studies, provide powerful clues about the nature and
timescale for internal angular momentum transport in stars. We focus on two cases: the metal poor
red giant KIC 7341231 (``Otto'') and intermediate mass core helium burning stars. For both we examine limiting case studies for angular momentum coupling between cores and envelopes under the assumption of rigid rotation on the main sequence. We discuss the expected pattern of core rotation as a function of mass and radius.
In the case of Otto, strong post-main-sequence coupling is ruled out and the measured core rotation rate is in the
range of 23 to 33 times the surface value expected from standard spin down models. The minimum coupling
time scale (.17  to .45 Gyr) is significantly longer than that inferred for young open cluster
stars. This implies ineffective internal angular momentum transport in early first ascent giants. By contrast, the core rotation rates of evolved
secondary clump stars are found to be consistent with strong coupling given their rapid main sequence
rotation. An extrapolation to the white dwarf regime predicts rotation periods between 330 and .0052 days depending on mass and decoupling time. We identify two key ingredients that explain these features: the presence of a convective core and inefficient angular momentum transport in the presence of larger mean molecular weight gradients. Observational tests that can disentangle these effects are discussed.

\end{abstract}

\section{Introduction}

Stars rotate; this basic fact is neglected in the standard theory of stellar evolution. This omission can be traced to two intertwined causes: angular momentum evolution theory is complex and empirical constraints of internal rotation have been scarce. In this paper we explore the consequences of recent detections of core rotation rates in red giants for internal angular momentum transport in stars, focusing on limiting case studies.

Until now we have relied on three diagnostics of internal rotation in low mass stars. As deduced from helioseismology, solar rotation is consistent with strong coupling between the radiative core and convective envelope mediated by a shear, or tachocline, layer at the convection zone base \citep{Schou1998}. Differential rotation has been claimed in the deep solar core based on g-mode data \citep{Garcia2007}, but the detection of such modes in the Sun remains controversial \citep[see][] {Appourchaux2010}. The spin down of young open cluster stars provides a more stringent test of the timescale for internal angular momentum transport. Low mass stars experience angular momentum loss from magnetized winds \citep{Weber1967}; the spindown rate becomes large when they arrive on the main sequence. In response, the convection zone initially spins down relative to the core until the angular momentum in the core can respond. The inferred core-envelope coupling timescale, of order tens to hundreds of Myr \citep[for example,][] {Keppens1995, Krishnamurthi1997, Bouvier1997, Denissenkov2010, Spada2011}, is consistent with solar constraints.

  We have indirect evidence for strong differential rotation based on the survival of rapid rotation in evolved blue core He-burning halo stars \citep{Peterson1983, Behr2003} in the presence of slow rotation at the main sequence turnoff and mass loss on the red giant branch \citep{Pinsonneault1991}. Slower rotation in the hottest halo giants, which experience strong gravitational settling (Behr 2003), is evidence for a suppression of angular momentum transport in the presence of strong near-surface mean molecular weight gradients \citep{Sills2000}. 

Structural evolution in red giants generates strong differential rotation with depth. Their core rotation is therefore a sensitive diagnostic for the timescale of internal angular momentum transport. Kepler has detected core rotation in low mass red giants \citep{Beck2012, Deheuvels2012}.  This surprising result is possible because the structure of red giants permits coupling between g modes excited in the core and p modes excited in the envelope for modes of spherical harmonic degree $\ell\ge$1 \citep{Dupret2009}. Modes with $\ell=0$, by contrast, are pure p-modes. Rapid core rotation then manifests itself as rotational splitting seen only in core-influenced modes. The phenomenon of differential rotation appears to be general \citep{Mosser2012}, which may permit a systematic exploration of the internal transport of angular momentum in stars.

Such data is essential because the theoretical problem is challenging. There are three broad categories of physical mechanisms that could transport angular momentum in stellar interiors: hydrodynamic, wave-driven, and magnetic \citep{Pinsonneault1997}. Hydrodynamic mechanisms can be effective in the rapid rotation regime, but the time scales for angular momentum transport become too long to be effective in the Sun \citep{Pinsonneault1989} or in evolved red giants \citep{Palacios2005, Chaname2005}. Additional processes are clearly required. Magnetic fields could be extremely effective for internal angular momentum transport \citep{Eggenberger2005}, although the precise impact can depend sensitively on the morphology \citep{Charbonneau1993}. Intermediate time scales for wave-driven transport are predicted, but  significant theoretical uncertainties still exist \citep{Talon2002}. There is no current consensus on which effect predominates.

Radically different coupling timescales are therefore permitted in principle. The expected evolution of high and low mass stars is also very different, and the mapping between core and surface is not intuitive. In this paper, we therefore consider limiting case scenarios for angular momentum evolution ranging from rigid rotation at all times to local conservation of angular momentum in the core. We explore intermediate coupling cases to gain insight into the relevant timescales. We compare these models with data in the low and intermediate mass regimes. We demonstrate that even at this abstract level, asteroseismology places powerful constraints on the permitted classes of models and discuss regimes where empirical data would be valuable. We describe our data and methods in Section 2, present our results in Section 3, and discuss them in Section 4.

\section{Methods}

\subsection{Stellar Models}
Our models are made using the Yale Rotating Evolution Code(YREC) \citep[see][ for a discussion of the input physics]{vanSaders2012}. An initial helium Y=0.270882, metallicity Z=0.018214, and mixing length $\alpha = 1.932706$ were chosen so that a solar model reproduces the solar luminosity ($3.827 \times 10^{33}$ ergs s$^{-1}$) and radius ($6.958 \times 10^{10}$ cm) at the solar age (4.57 Gyrs).

The range of theoretical possibilities discussed above can be captured by limiting case scenarios. For a given initial condition and loss law, rigid rotation enforced at all times (``Maximal Coupling'') sets a lower bound on the permitted core rotation. Local conservation of angular momentum at all times in radiative zones defines an upper envelope (``No Coupling''). We also explore the effects of enforcing a solar-like rotation paradigm (rigid rotation) on the main sequence but decoupling the radiative regions in the models during the post-main-sequence evolution (``No Coupling Post-Main-Sequence''). This approach allows us to infer core rotation rates requiring differential rotation, and set bounds on the evolutionary state where it emerged. We enforce depth-independent rotation in convection zones; differential convection zone rotation would result in weaker bounds.

We include angular momentum loss in all models with T$_{eff} <9500$ K. Angular momentum loss is thus not included in intermediate mass main sequence stars, which do not experience magnetized solar-like winds. Post-main-sequence angular momentum loss had a 20\% effect on the predicted surface rotation rate of the low mass case and a less than 1\% effect on the intermediate mass case. Main sequence loss produces extremely significant effects. We model angular momentum loss using a modified \citet{Kawaler1988} prescription: $\frac{dJ}{dt}=-K\omega^3(\frac{R}{R_\sun})^{0.5}(\frac{M}{M_\sun})^{-0.5}$ up to a Rossby scaled critical rotation rate of $\omega_{crit}=3 \times 10^{-5} \frac{\tau_\sun}{\tau_{star}}$. For more rapid rotation  $\frac{dJ}{dt}=-K\omega\omega_{crit}^2(\frac{R}{R_\sun})^{0.5}(\frac{M}{M_\sun})^{-0.5}$, with $K=2.65 \times 10^{47}$ s \citep{Sills2000, Krishnamurthi1997}. 

\subsection{Model Initial Conditions}
There is a radical difference between the expected behavior of high and low mass giants because of their very different initial conditions. Low mass stars have radiative main sequence cores and experience strong angular momentum loss. Their predicted post-main-sequence surface rotation rates are therefore low and severe mass loss erases the memory of the initial conditions. Stars around 2.5 M$_\sun$ experience little main sequence angular momentum or mass loss. Unlike the low mass case, these stars leave the main sequence with both a high mean rotation and a substantial range in rotation rates. Such stars also do not experience a  helium flash or significant angular momentum loss as they cross from the main sequence to the secondary red clump.

We have core rotation rates and surface gravities for both Otto \citep{Deheuvels2012} and secondary clump stars \citep{Mosser2012}. Otto has M$=0.856$ M$_\sun$, [Fe/H]$=-1$, $[\alpha/Fe] = +0.2$, Y=.253392, and $\alpha$=1.93, log(g)$=3.520 \pm 0.007$, T$_{eff}=5363 \pm 118 $K, $\Omega_c=710 \pm 51$ nHz, and $\Omega_s<150 \pm 19$ nHz \citep{Deheuvels2012}. For our investigation of intermediate mass stars, we use recent results for stars greater than 2 M$_\sun$ from \citet{Mosser2012}.

\subsubsection{Low Mass Initial Rotation}
In the absence of direct metal-poor data, we adopt Population I initial conditions: P$_{init}$ = 8 days and a disk locking period of 3 Myrs \citep{Tinker2002}. Our results are insensitive to the choice of initial conditions because of severe angular momentum loss. To determine the age at which our models should match the observed parameters, we compared models to the asteroseismic mean density $(\bar{\rho}=0.06457$ g cm$^{-3})$, as given by $\Delta\nu$.

There are degeneracies between the seismic mass and the adapted metallicity. We therefore consider two alternate initial conditions from \citet{Deheuvels2012}, a lower metallicity case (M$=0.80$ M$_\sun$, $Y=0.250922$, [Fe/H]$=-1.25$) and a higher metallicity case (M$=0.90$M$_\sun$, $Y=0.257834$, [Fe/H]$=-0.75$). 

\subsubsection{Intermediate Mass Initial Rotation}
To examine the rotational evolution of intermediate mass stars, we evolved a 2.5 M$_\sun$, solar metallicity model to approximate the secondary clump giants analyzed by \citet{Mosser2012}. Because there is insignificant angular momentum loss on the main sequence, the observed spread in main sequence rotation rates (from 50-300  km s$^{-1}$) \citep{Dufton2006} will carry over into the post-main-sequence. We construct a model with a characteristic main sequence surface rotation rate (150 km s$^{-1}$)  using a disk rotation period of 2.97 days and a disk locking time of 3 Myrs. We also explore the effects of assuming a slow rotator (main sequence rotation rate of 50 km s$^{-1}$), and a fast rotator (main sequence rotation rate of 300 km s$^{-1}$).

\section{Results}
Different scenarios for core-envelope coupling predict radically different core rotation rates (Figure 1). We begin by reviewing the physical causes of these differences (\S 3.1) and then explore the consequences for the minimum coupling timescale (\S 3.2).

\subsection{Overall Predicted Behavior}
In the decoupled case, the dramatic core contraction in the giant phase dictates the change in rotation rate. If the core is strongly coupled to the envelope, in contrast, the four orders of magnitude increase in the moment of inertia is the dominant effect. Core rotation rates from seconds to centuries are therefore possible in red giants. In the intermediate case (a star which rotates as a solid body on the main sequence but is decoupled post-main-sequence) the rotation rate asymptotes to P$_{rot} \sim .2$  and P$_{rot} \sim 0.005$ days after the dredge up for the low and intermediate mass cases, respectively. Uncertainties in model parameters (mass and metallicity in the low mass case and main sequence rotation rate in the intermediate mass case) produce differences in the evolved rotation rates that are small compared to the difference produced by varying the coupling assumptions.

The post-main-sequence structural evolution of low and intermediate mass stars are distinctly different and the main events which might affect angular momentum evolution are labeled in Figure 1. As soon as stars leave the main sequence their surface convection zones (SCZ) deepen. As the SCZ penetrates into regions where nuclear burning took place, a large mean molecular weight gradient develops at the base of the convection zone. Stars also experience a drop in luminosity and radius when the hydrogen burning shell crosses the discontinuity in the mean molecular weight left over from the dredge up, producing a red giant branch bump. The next major structural change occurs at the helium flash, where the core has contracted enough that helium fusion begins. Otto is past the onset of the dredge-up and before the red giant branch bump. Its rotation rate lies squarely between the strongly coupled track and the coupled main sequence, decoupled post-main-sequence track, indicating that it decouples at some point during its post-main-sequence evolution. 

After intermediate mass stars leave the main sequence they rapidly cross the Hertzsprung gap. Helium fusion is ignited in a non-degenerate core, after which the star smoothly contracts back to and settles on the red clump. Interestingly, the \citet{Mosser2012} sample of secondary clump stars have core rotation rates consistent with both the mean and spread of our strongly coupled models. 

\begin{figure*}[h]
 \centering
 \subfigure{
\includegraphics[width=0.80\textwidth, clip=true, trim=1.5cm 0.8cm 0.5cm 1.2cm]   
         {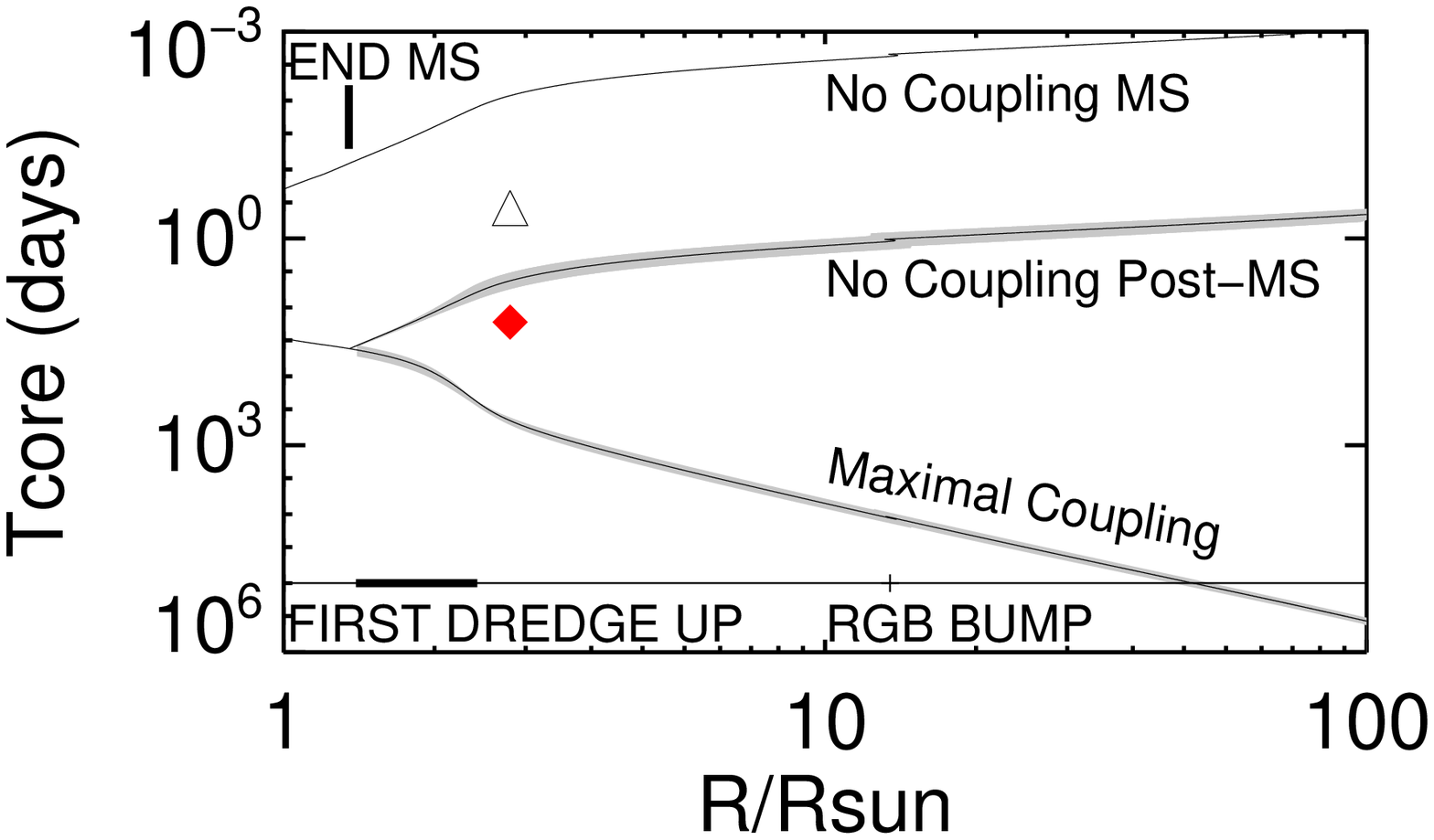}
 }
 \subfigure{\rlap{ \raisebox{2.5cm}{\includegraphics[height=2.4cm,clip=true, trim=-10.cm 0.8cm 0.5cm 1.4cm]
              {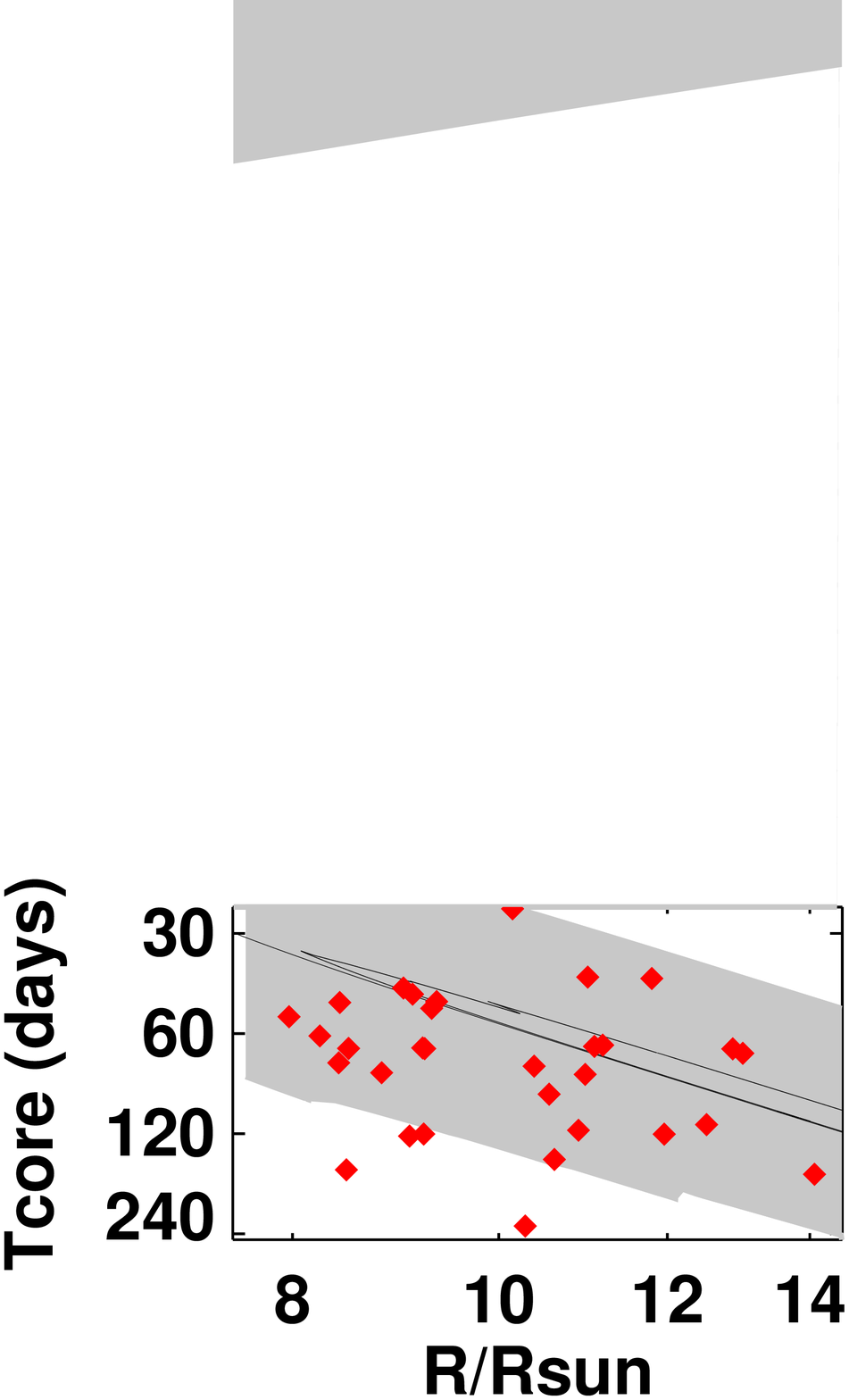}}}
\includegraphics[width=0.80\textwidth, clip=true, trim=1.5cm 0.8cm 0.5cm 1.2cm]   
         {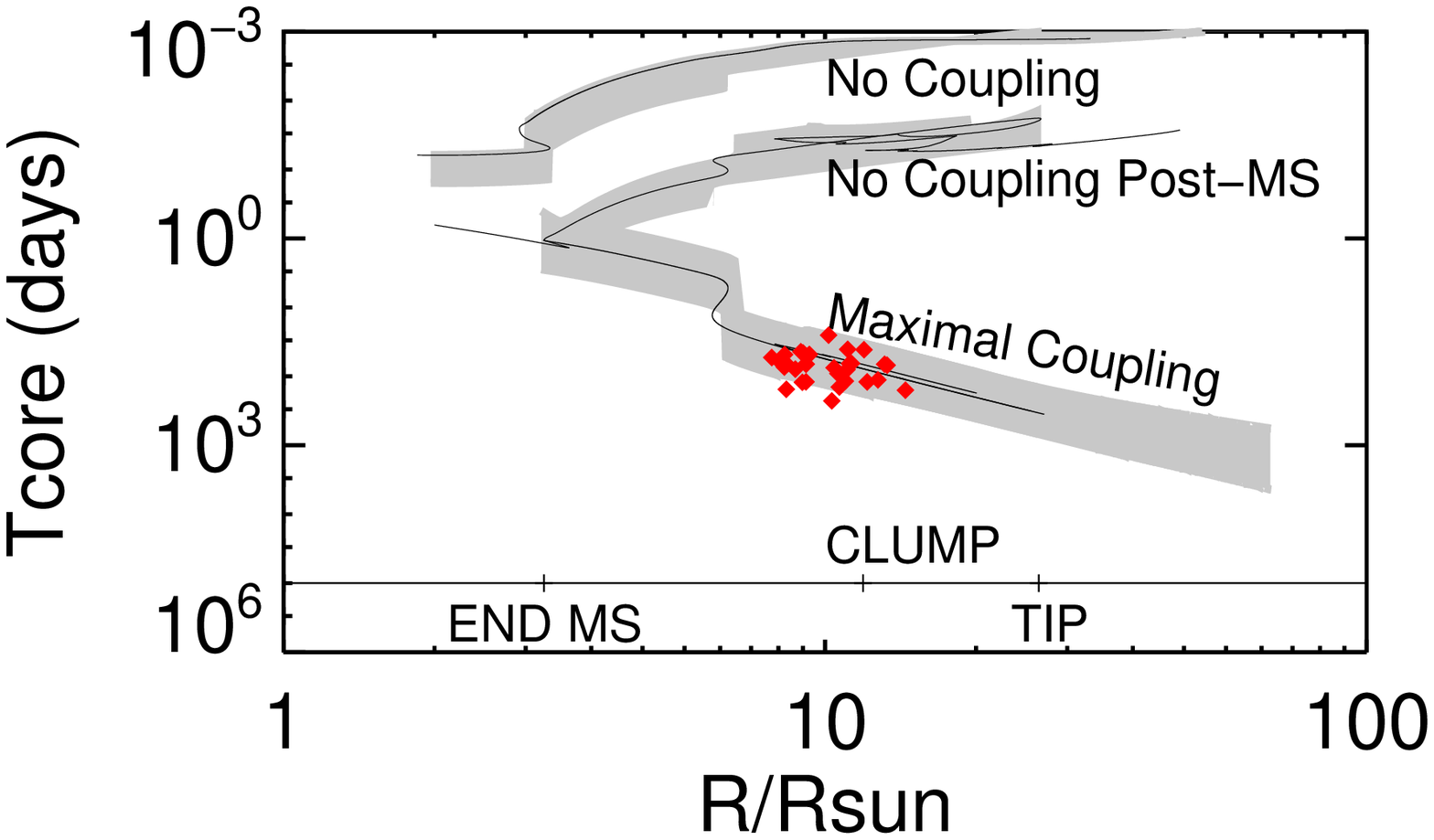}

 }
\caption {The top plot represents models run for the low mass case, representing Otto, while the lower plot represents the intermediate mass ($2.5M_\sun$) case. From top to bottom, the three broad bands on each plot indicate the predicted core rotation rates given no internal angular momentum transport; solid body rotation on the main sequence and no coupling post-main-sequence; and solid body rotation for the entire run. Grey bands indicate uncertainties due to mass and metallicity(top plot) and the predicted range of main sequence rotation rates (bottom plot). The triangle indicates the model by \citet{Ceillier2012}. Red diamonds represent the Deheuvels best fit model for Otto (top) and all observations above 2 M$_\sun$ from \citet{Mosser2012}(bottom). The inset bottom panel figure zooms in on the red clump. }

\end{figure*}

\begin{figure*}[h]
 \centering
\subfigure{
\includegraphics[width=0.85\textwidth, clip=true, trim=1.2cm 0.8cm 0.5cm 1.2cm] 
        {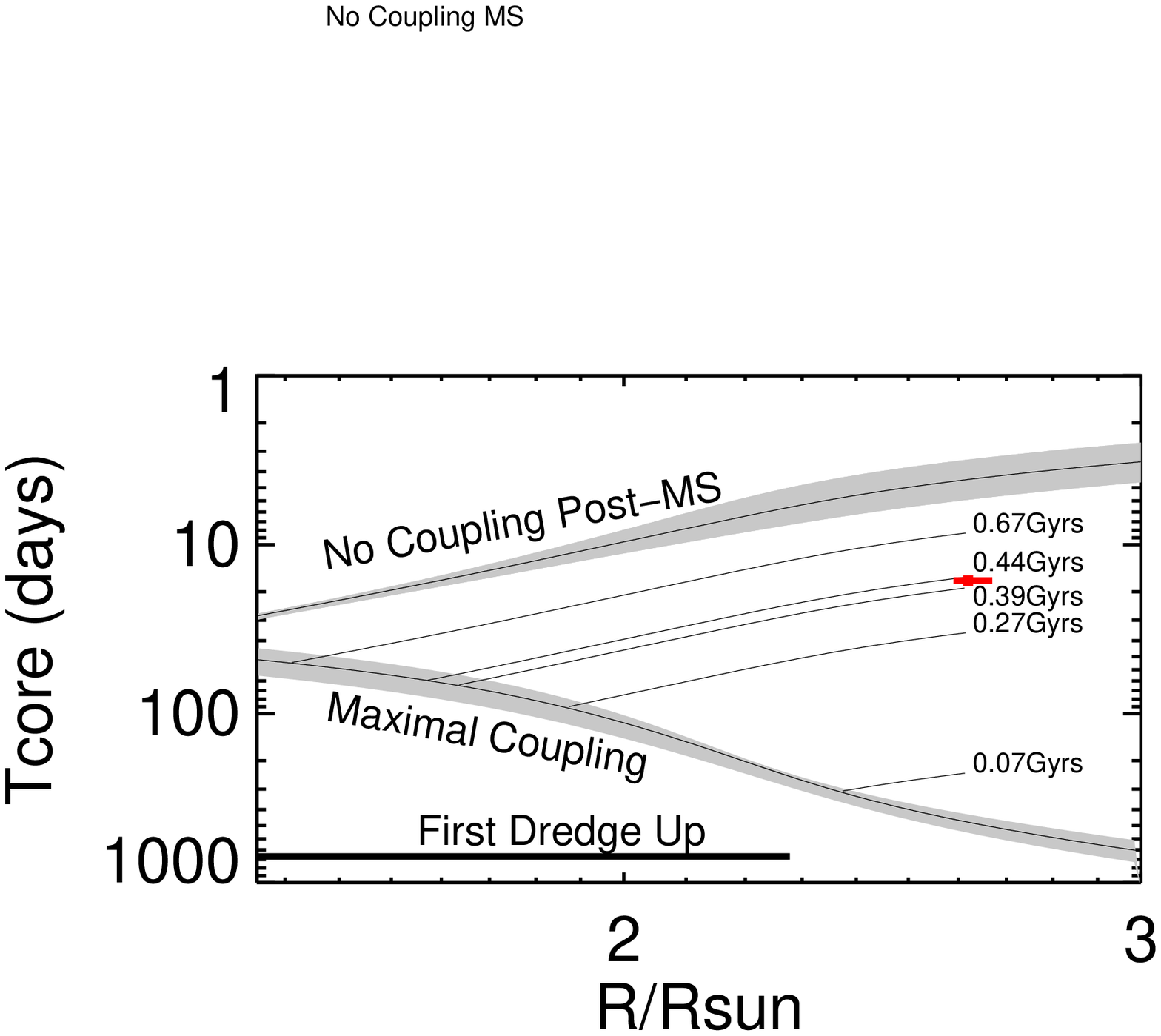}
}

\subfigure{
\includegraphics[width=0.85\textwidth, clip=true,   trim=1.2cm 0.8cm 0.5cm 1.2cm]  
      {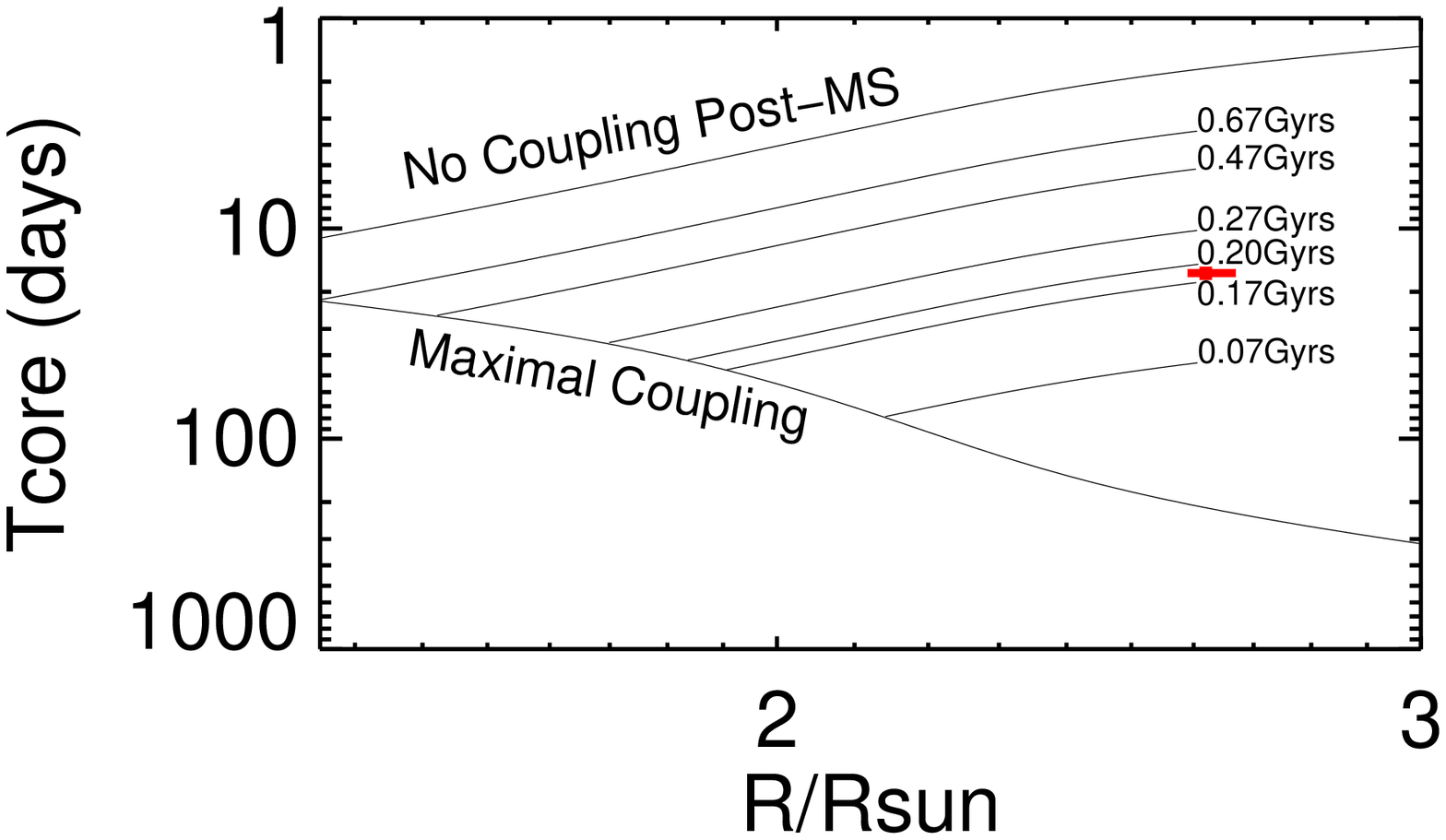}
}

\caption {The top plot represents the best fit low mass (Otto) model run using  pre-main-sequence rotation data as a starting point. In the bottom plot are models run using the main sequence turnoff rotation rates at the spectroscopic upper limit. Tracks between the solid body (bottom) and no post-main-sequence coupling (top) tracks explore the effects of relaxing solid body rotation at different points on the subgiant branch.. A minimum decoupling timescale of .45 Gyrs is required to reproduce the observations for Otto using reasonable pre-main-sequence rotation assumptions; using the main sequence spectroscopic upper limits decreases the minimum coupling time to .17 Gyrs. }
\end{figure*} 

\subsection{Minimum Coupling Timescales: Low Mass Stars}

Our models for Otto require decoupling after the main sequence. To determine the minimum coupling timescale, we evolved models with solid body rotation enforced up to a series of specific times. These final states are then used as starting points for models with local conservation of angular momentum in radiative regions, which are then evolved to the mean density of Otto. We calculate a minimum decoupling timescale of $0.415 \pm 0.03$ Gyrs. Decoupling therefore occurs before the maximum mean molecular weight gradient develops at the base of the convection zone during the first dredge up.  These models predict a surface that rotates 23 to 33 times slower than the core, a prediction which could be used to further constrain the internal rotation profile of the star. This timescale is significantly longer than published early main sequence timescales ($55 \pm 25$ Myrs and 130 Myrs from \citet{Denissenkov2010} and \citet{Spada2011}, respectively). 

Halo stars are sharp-lined objects, but our predicted main sequence core rotation rate (1.2 km s$^{-1}$) is below the empirical limit (4 km s$^{-1}$)\citep{Lucatello2003}. To get an even stricter bound on the minimum decoupling timescale, we consider a model that leaves the main sequence at this limit. These very conservative assumptions yield a shorter minimum coupling timescale of $0.183 \pm 0.02$ Gyrs and a surface which rotates 10.9 to 12.5 times slower than the core.

\section{Discussion}
We find qualitative differences in the angular momentum coupling strengths of low and intermediate mass stars. While observations indicate stars in both mass ranges need to be strongly coupled on the main sequence, low mass stars appear to decouple around the epoch of the first dredge up, while intermediate mass stars are strongly coupled on the red clump. Unfortunately, data is not yet available for intermediate mass stars crossing to the clump, as crossing times are short. 

Our models can be compared with the recent calculations of \citet{Ceillier2012, Ceillier2013}. The 2012 analysis predicted a rotation rate an order of magnitude faster than Otto's observed rate. Our limiting case models indicate that this is due to a lack of main sequence coupling, rather than to differences in post-main-sequence evolution \citep[see][for a related discussion]{Ceillier2013}. While the mechanism enforcing the strong coupling  measured in main sequence stars is not yet fully understood, giant branch core rotation rates published to date do no require differential rotation on the main sequence.

In the following sections, we suggest two possible structural changes which could be affecting the coupling strength. We describe observational tests of these hypotheses (\S 4.1 and 4.2) and discuss the implications for white dwarf rotation rates (\S 4.3). We close with a discussion of important potential selection effects (\S 4.4).

\subsection{Transport Inhibition by Mean Molecular Weight Gradients}
Low mass stars appear to decouple during the first dredge up as strong mean molecular weight gradients develop at the base of the convection zone. A natural explanation is that these gradients inhibit angular momentum transport, as in horizontal branch globular cluster stars, where a break in the measured surface rotation coincides with the onset of gravitational settling \citep{Sills2000}. 

This theory predicts that decoupling sets in at the first dredge up in all stars, proportional to the strength of their mean molecular weight gradient at the base of the convection zone. A clear observational test of such models would be a measurement of recoupling after the red giant branch bump, when these gradients disappear. One interesting puzzle is that core mean molecular weight gradients, as opposed to envelope ones, do not appear to have a direct observational signature yet. An absence of recoupling on the upper red giant branch would be powerful evidence for their impact.

\subsection{Transport Enhancement by Convective Cores}

Intermediate mass stars are strongly coupled on the secondary clump; this could be related to the presence of a convective core. The detection of rapid horizontal branch rotation in globular cluster stars requires strong recoupling on the horizontal branch \citep{Sills2000}, consistent with this data. If core-driven gravity waves are significant, we expect strong coupling to be a general feature of stars with convective cores. Weaker coupling would then set in for shell burning stars and stars with radiative cores, including intermediate mass stars observed in the Hertzsprung Gap. A low initial surface rotation rate would be expected for stars that travel quickly from the tip to the core helium burning phase.  If gravity waves are the coupling mechanism, we could therefore compute the coupling timescale by examining the spin up of stars on the red clump.

\subsection{Consequences for White Dwarf Rotation Rates}
We noted in section 3.1 that a decoupled low mass star's rotation rate asymptotes to P$_{rot} \sim 0.2$ days, a number intriguingly similar to typical white dwarf rotation rates \citep{Charpinet2009}. Assuming evolution does not alter the scaling between core rotation rate and central density, we estimate white dwarf rotation periods. Decoupling at the end of the low mass main sequence gives a period of 0.20 days, while decoupling at a radius of 10 R$_\sun$ (approximating the clump) predicts P$_{rot} \sim 330$ days. In the intermediate mass case, decoupling at the end of the main sequence gives a rotation period of .0052 days, while decoupling on the clump yields a rotation period of 3.3 days. Measured white dwarf rotation rates then imply that low mass stars remain mostly decoupled through their post-main-sequence evolution, while intermediate mass stars decouple after the secondary clump.

\subsection{Possible Selection Effects}
 \citet{Deheuvels2012} indicate that core rotation rates are robust but internal profiles are not. While that paper used a full inversion, such analysis was not feasible for ensemble asteroseismology \citep[for example]{Mosser2012}. Additionally, of the 1400 red giants analyzed by seismologists, only about 700 have measured evolutionary stages, and only about 300 of those have identifiable mixed modes and rotational splittings \citep{Bedding2011, Mosser2012, Mosser2012a}. Depressed mixed modes due to weak coupling, inclination effects and confusion between mixed mode and rotational splittings likely contribute to nondetections \citep{Mosser2012c}. Measurements of matched core and surface rotation rates, core rotation rates of low mass stars past the red giant branch bump, intermediate mass stars in the Hertzsprung gap, as well rotation rates of distincly young or old clump giants are not yet available and are required to address potential selection effects.

\section{Conclusion}
We are able to use models to place reasonable limits on the expected core rotation rates for both low (0.85 $M_\sun$) and intermediate (2.5 $M_\sun$) mass stars. To match observations, low mass stars require significant decoupling between the core and envelope that sets in around the first dredge up, while intermediate mass stars are strongly coupled on the secondary clump. We hypothesize that this difference is due to either effective angular momentum transport from convective cores in intermediate mass stars, which allows additional angular momentum to be transfered through gravity waves, or the development of a mean molecular weight gradient at the base of the convection zone in low mass subgiants, which inhibits angular momentum transport between the core and envelope. Further progress will require complementary theoretical and observation work. For theory, neither weakly coupled hydrodynamic models nor strongly coupled magnetic ones are compatible with the data.  Key observational questions include the robustness of the core rotation rates, observational selection effects and surface rotation measurements.  Key data includes detecting the epoch of decoupling in subgiants and the presence or absence of recoupling above the red giant branch bump.  Measurements of core rotation as a function of log(g) on the red clump could constrain the coupling timescale for convective cores.  Finally, strong mass trends in main sequence rotation (van Saders \& Pinsonneault 2013, in prep) will be important for interpreting core rotation data.

\begin{acknowledgements}
We would like to thank Benoit Mosser for helpful discussions and for providing the core rotation data for the secondary clump stars.  This work was supported by NASA  grant NNX11AE04G.
\end{acknowledgements}

\end{document}